\documentclass{elsart}
\usepackage{amsmath}
\usepackage{axodraw}
\usepackage{ifthen}

\def\xdim{40}   
\def\ydim{40}
\def\xoff{0}    
\def\yoff{0}
\def\dash{2}    
\def\vtxsz{1}   
\def\heavydotsz{4} 

\SetScale{1}
\SetWidth{0.5}

\newcommand{\LA}[1]{\ifthenelse{\equal{#1}{2}}{\CArc(20,20)(19.25,0,90)\CArc(20,20)(20.75,0,90)}{}\ifthenelse{\equal{#1}{1}}{\CArc(20,20)(20,0,90)}{}\ifthenelse{\equal{#1}{0}}{\DashCArc(20,20)(20,0,90){\dash}}{}}
\newcommand{\LB}[1]{\ifthenelse{\equal{#1}{1}}{\CArc(40,40)(20,180,270)}{\DashCArc(40,40)(20,180,270){\dash}}}
\newcommand{\LC}[1]{\ifthenelse{\equal{#1}{1}}{\CArc(20,20)(20,180,270)}{\DashCArc(20,20)(20,180,270){\dash}}}
\newcommand{\LD}[1]{\ifthenelse{\equal{#1}{1}}{\CArc(20,20)(20,90,180)}{\DashCArc(20,20)(20,90,180){\dash}}}
\newcommand{\LE}[1]{\ifthenelse{\equal{#1}{1}}{\Line(20,0)(20,20)}{\DashLine(20,0)(20,20){\dash}}}
\newcommand{\LF}[1]{\ifthenelse{\equal{#1}{1}}{\Line(20,20)(20,40)}{\DashLine(20,20)(20,40){\dash}}}
\newcommand{\LG}[1]{\ifthenelse{\equal{#1}{1}}{\Line(20,20)(0,20)}{\DashLine(20,20)(0,20){\dash}}}
\newcommand{\LH}[1]{\ifthenelse{\equal{#1}{k}}{\PhotonArc(20,20)(20,270,360){1}{9}}{}\ifthenelse{\equal{#1}{3}}{\CArc(20,20)(19,270,360)\CArc(20,20)(20,270,360)\CArc(20,20)(21,270,360)}{}\ifthenelse{\equal{#1}{2}}{\CArc(20,20)(19.25,270,360)\CArc(20,20)(20.75,270,360)}{}\ifthenelse{\equal{#1}{1}}{\CArc(20,20)(20,270,360)}{}\ifthenelse{\equal{#1}{0}}{\DashCArc(20,20)(20,270,360){\dash}}{}}

\newcommand{\KA}{\CArc(20,35)(5,0,360)}
\newcommand{\KC}[1]{\ifthenelse{\equal{#1}{1}}{\CArc(20,15)(15,180,270)}{\DashCArc(20,15)(15,180,270){\dash}}}
\newcommand{\KD}[1]{\ifthenelse{\equal{#1}{1}}{\CArc(20,15)(15,90,180)}{\DashCArc(20,15)(15,90,180){\dash}}}
\newcommand{\KE}[1]{\ifthenelse{\equal{#1}{1}}{\Line(20,0)(20,15)}{\DashLine(20,0)(20,15){\dash}}}
\newcommand{\KF}[1]{\ifthenelse{\equal{#1}{1}}{\Line(20,15)(20,30)}{\DashLine(20,15)(20,30){\dash}}}
\newcommand{\KG}[1]{\ifthenelse{\equal{#1}{1}}{\Line(20,15)(5,15)}{\DashLine(20,15)(5,15){\dash}}}
\newcommand{\KH}[1]{\ifthenelse{\equal{#1}{3}}{\CArc(20,15)(14,-90,90)\CArc(20,15)(15,-90,90)\CArc(20,15)(16,-90,90)}{}\ifthenelse{\equal{#1}{2}}{\CArc(20,15)(14.25,-90,90)\CArc(20,15)(15.75,-90,90)}{}\ifthenelse{\equal{#1}{1}}{\CArc(20,15)(15,-90,90)}{}\ifthenelse{\equal{#1}{0}}{\DashCArc(20,15)(15,-90,90){\dash}}{}}

\newcommand{\SeTopo}[8]{\LA{#1}\LB{#2}\LC{#3}\LD{#4}\LE{#5}\LF{#6}\LG{#7}\LH{#8}\Vertex(0,20){\vtxsz}\Vertex(20,0){\vtxsz}\Vertex(20,20){\vtxsz}\Vertex(20,40){\vtxsz}\Vertex(40,20){\vtxsz}}
\newcommand{\TeTopo}[6]{\KA\KC{#1}\KD{#2}\KE{#3}\KF{#4}\KG{#5}\KH{#6}\Vertex(5,15){\vtxsz}\Vertex(20,0){\vtxsz}\Vertex(20,15){\vtxsz}\Vertex(20,30){\vtxsz}}

\newcommand{\Topo}[1]{
\ifthenelse{\equal{#1}{SA}}{\SeTopo{0}{0}{1}{1}{1}{1}{0}{k}}{} 
\ifthenelse{\equal{#1}{SB}}{\SeTopo{1}{0}{0}{1}{1}{0}{1}{k}}{} 
\ifthenelse{\equal{#1}{SC}}{\SeTopo{1}{1}{1}{1}{1}{1}{0}{k}}{} 
\ifthenelse{\equal{#1}{SD}}{\SeTopo{0}{0}{1}{0}{1}{0}{1}{k}}{} 
\ifthenelse{\equal{#1}{SE}}{\SeTopo{1}{0}{1}{1}{0}{0}{0}{k}}{} 
\ifthenelse{\equal{#1}{SF}}{\SeTopo{1}{1}{1}{0}{1}{0}{1}{k}}{} 
\ifthenelse{\equal{#1}{SFo}}{\SeTopo{1}{1}{1}{0}{1}{0}{1}{0}}{} 
\ifthenelse{\equal{#1}{SG}}{\SeTopo{1}{1}{0}{0}{0}{0}{0}{k}}{} 
\ifthenelse{\equal{#1}{WB}}{\SeTopo{1}{0}{0}{1}{1}{0}{1}{0}}{}
\ifthenelse{\equal{#1}{WC}}{\SeTopo{1}{1}{1}{1}{1}{1}{0}{2}}{}
\ifthenelse{\equal{#1}{WE}}{\SeTopo{1}{0}{1}{1}{0}{0}{0}{0}}{}
\ifthenelse{\equal{#1}{WF}}{\SeTopo{1}{1}{1}{0}{1}{0}{1}{2}}{}
\ifthenelse{\equal{#1}{WG}}{\SeTopo{1}{1}{0}{0}{0}{0}{0}{2}}{}
\ifthenelse{\equal{#1}{XB}}{\SeTopo{1}{0}{0}{1}{1}{0}{1}{3}}{}
\ifthenelse{\equal{#1}{TB}}{\TeTopo{0}{1}{1}{0}{1}{0}}{}
\ifthenelse{\equal{#1}{TC}}{\TeTopo{1}{1}{1}{1}{0}{0}}{}
\ifthenelse{\equal{#1}{TE}}{\TeTopo{1}{1}{0}{0}{0}{0}}{}
\ifthenelse{\equal{#1}{TF}}{\TeTopo{1}{0}{1}{0}{1}{0}}{}
\ifthenelse{\equal{#1}{TG}}{\TeTopo{0}{0}{0}{0}{0}{0}}{}
\ifthenelse{\equal{#1}{UB}}{\TeTopo{0}{1}{1}{0}{1}{1}}{}
\ifthenelse{\equal{#1}{UC}}{\TeTopo{1}{1}{1}{1}{0}{2}}{}
\ifthenelse{\equal{#1}{UE}}{\TeTopo{1}{1}{0}{0}{0}{1}}{}
\ifthenelse{\equal{#1}{UF}}{\TeTopo{1}{0}{1}{0}{1}{2}}{}
\ifthenelse{\equal{#1}{UG}}{\TeTopo{0}{0}{0}{0}{0}{2}}{}
\ifthenelse{\equal{#1}{VB}}{\TeTopo{0}{1}{1}{0}{1}{3}}{}
}

\newcommand{\GenericSeTopo}{
  \raisebox{-9mm}{\begin{picture}(60,60)(\xoff,\yoff)\SetScale{1.5}
    \LA{1}\LB{1}\LC{1}\LD{1}\LE{1}\LF{1}\LH{1}\put(44.68,8.25){\vector(1,1){10}}
    \Vertex(20,0){\vtxsz}\Vertex(20,40){\vtxsz}\Vertex(40,20){\vtxsz}
    \Text(15,30)[]{\small{$R^{\mu_1\dots}$}}\Text(45,45)[]{\small{~$S_{\mu_1..}$}}\Text(53,10)[lt]{\small{$a_0,m_0$}}\Text(49,14)[rb]{\small{$q$}}
  \SetScale{1}\end{picture}}
}
\newcommand{\FireSeTopo}[1]{
  \raisebox{-6mm}{\begin{picture}(40,40)(\xoff,\yoff)
  \LA{2}\LC{1}\LD{0}\LE{1}\LF{0}\LG{1}\LH{0}
  \Vertex(0,20){\vtxsz}\Vertex(20,0){\vtxsz}\Vertex(20,20){\vtxsz}\Vertex(20,40){\vtxsz}\Vertex(40,20){\vtxsz}
  \ifthenelse{\equal{#1}{1}}{\Vertex(40,20){\heavydotsz}}{}
  \end{picture}}
}

\newcommand{\TopoA}[1]{\raisebox{-6mm}{\begin{picture}(\xdim,\ydim)(\xoff,\yoff)\Topo{#1}\end{picture}}}
\newcommand{\TopoB}[2]{\raisebox{-6mm}{$\underset{#2}{\begin{picture}(\xdim,\ydim)(\xoff,\yoff)\Topo{#1}\end{picture}}$}}

\def\labelszC{\scriptsize}
\newcommand{\TopoC}[1]{\raisebox{-6mm}{\begin{picture}(\xdim,\ydim)(\xoff,\yoff)\Topo{#1}
  \Text(36,8)[lt]{\labelszC{$a_0$}}\Text(36,32)[lb]{\labelszC{$a_1$}}\Text(30,28)[]{\labelszC{$a_2$}}\Text(4,8)[rt]{\labelszC{$a_3$}}\Text(4,32)[rb]{\labelszC{$a_4$}}\Text(20,10)[r]{\labelszC{$a_5$}}\Text(20,30)[r]{\labelszC{$a_6$}}\Text(11,19)[t]{\labelszC{$a_7$}}
\end{picture}}}

\def\labelszD{\scriptsize}
\newcommand{\TopoD}[2]{\raisebox{-6mm}{\begin{picture}(50,40)(-5,\yoff)\Topo{#1}
  \Text(37,8)[lt]{\labelszD{$r_#2$}}\Text(6,8)[rt]{\labelszD{$b_1$}}\Text(6,35)[rb]{\labelszD{$b_2$}}\Text(21,10)[l]{\labelszD{$b_3$}}\Text(20,30)[r]{\labelszD{$b_4$}}\Text(11,19)[t]{\labelszD{$b_5$}}
\end{picture}}}

\def\labelszE{\tiny}
\newcommand{\TopoE}[2]{\raisebox{-6mm}{\begin{picture}(50,40)(-5,\yoff)\Topo{#1}
  \Text(36,15)[l]{\labelszE{$r_#2$}}\Text(8,7.5)[rt]{\labelszE{$b_1$}}\Text(8,25)[rb]{\labelszE{$b_2$}}\Text(21,7.5)[l]{\labelszE{$b_3$}}\Text(21,22.5)[l]{\labelszE{$b_4$}}\Text(13,14)[t]{\labelszE{$b_5$}}
\end{picture}}}

\allowdisplaybreaks

\def\vtxsz{1.5}     
\def\heavydotsz{5}  

\begin{document}
\begin{frontmatter}
\begin{flushleft}
TTP09-18\\
SFB/CPP-09-51\\
arXiv:0907.2117
\end{flushleft}
\title{Low energy moments of heavy quark current correlators at 
  four loops} 
\author[Karlsruhe]{A.~Maier}, 
\author[Karlsruhe]{P.~Maierh\"ofer}, 
\author[Karlsruhe]{P.~Marquard} and
\author[Moskau]{A.~V.~Smirnov}
\address[Karlsruhe]{Institut f\"ur Theoretische Teilchenphysik,
Universit\"at Karlsruhe, Karlsruhe~Institute~of~Technology~(KIT), 76128 Karlsruhe, Germany}
\address[Moskau]{
Scientific Research Computing Center of Moscow State University, Russia}
\begin{abstract}
  We describe several techniques for the calculation of multi-loop
  integrals and their application to heavy quark current correlators. As
  new results, we present the four-loop correction to the second and
  third physical moment in the low-energy expansions of vector,
  axial-vector and scalar quark current correlators. Using a Ward
  identity, we obtain the third and fourth moment for the
  pseudo-scalar correlator. We briefly discuss the impact of these results
  on the determination of the charm quark mass and the strong coupling
  constant using lattice simulations for the current correlators and of
  the charm- and bottom-quark mass from experimental data for
  $\sigma(e^+ e^- \to \mbox{hadrons})$.
\end{abstract}
\end{frontmatter}

\section{Introduction}
\label{sec:introduction}

One of the phenomenologically most interesting applications of heavy
quark current correlators is the determination of fundamental parameters
of QCD via sum rules. Low-energy moments of the vector correlator can be
compared to weighted integrals over the experimentally measured $R$-ratio,
$ R(s) = \sigma(e^+e^- \rightarrow \text{hadrons})/\sigma(e^+e^-
\rightarrow\mu^+\mu^-)$, to determine values for the charm- and
bottom-quark masses \cite{Shifman:1978by,Reinders:1984sr,Kuhn:2001dm,Kuhn:2007vp}. Recently ``data'' from lattice
simulations have been used instead of experimental results and indeed
this method has become a competitive way of extracting the charm-quark
mass and the strong coupling constant 
\cite{Allison:2008xk}. Although the correlators of all four
(axial-vector, vector, scalar and pseudo-scalar) currents can be used in
the lattice method, the most accurate predictions presented in
\cite{Allison:2008xk} were based on the pseudo-scalar correlator.

In order to obtain precise values for the quark
masses, it is mandatory to calculate higher order QCD
corrections. To match the experimental precision this means calculations
in four-loop approximation have to be performed. 
Improvements in both computer power and the techniques of
multi-loop calculations have lead to significant progress in recent years.

After the second moment at order $\alpha_s^3$ of the vector current
became available it was possible to construct a Pad\'e approximant of
the vacuum polarization function \cite{Hoang:2008qy}. This also leads to
a prediction for the higher moments. With the explicit calculation of
the third moment this prediction can be checked and it becomes possible
to further improve the Pad\'e
approximation \cite{Kiyo:2009}.

The three-loop, i.e. ${\cal O}(\alpha_s^2)$, corrections were evaluated
more than ten years ago in
\cite{Chetyrkin:1995ii,Chetyrkin:1996cf,Chetyrkin:1997mb} for vector,
axial-vector, scalar and pseudo-scalar currents in the low energy
expansion up to $(q^2)^8$. This calculation has employed the reduction
method proposed in \cite{Broadhurst:1991fi}. Recently, a different
approach based on the combination of the Laporta algorithm
\cite{Laporta:2001dd} with differential equations
\cite{Remiddi:1997ny,Caffo:1998du,Kotikov:1990kg} was used to calculate
these corrections for terms up to $(q^2)^{30}$ \cite{Boughezal:2006uu,
  Maier:2007yn}.

At four-loop order, for the vector correlator the first physical moment
proportional to $(q^2)^1$ was obtained in
\cite{Chetyrkin:2006xg,Boughezal:2006px}. The second moment of this
correlator was presented in \cite{Maier:2008he}, where, aside from the
Laporta algorithm, also techniques based on Sbases and special
treatment of internal self-energies were used for checks. These methods
will be explained in detail in the paper at hand. Using the method
mentioned in the previous paragraph, the thirty lowest moments of the
double-fermionic corrections at ${\cal O}(\alpha_s^3n_f^2)$ were
determined in \cite{Czakon:2007qi}. The part proportional to ${\cal
  O}(\alpha_s^nn_l^{(n-1)})$ is even known to all orders in perturbation
theory \cite{Grozin:2004ez}.

For the pseudo-scalar correlator, the first moment and second moment at
${\cal O}(\alpha_s^3)$ are given in \cite{Sturm:2008eb}, together with
the first moment of the axial-vector and the scalar correlator.

In this work, we present the calculation of the second and third moments
of the vector, axial-vector and scalar correlators and the third and
fourth moments of the pseudo-scalar correlator. In Section
\ref{sec:methods} we define our notation and explain the methods used
in the calculation. In Section \ref{sec:reduct-techn-integr} we present
a newly developed method for the efficient treatment of diagrams
containing self-energy insertions.  Section \ref{sec:results} contains
the results for the various currents and an update of the determination
of charm- and bottom-quark masses from experimental data and $\alpha_s$
from lattice calculations is given in Section 5. We conclude in Section
\ref{sec:conclusion}.
\section{Methods}
\label{sec:methods}
The polarization functions are defined by
 \begin{align}
   &(-q^2 g_{\mu\nu} + q_\mu q_\nu) \Pi^\delta (q^2) + q_\mu q_\nu
   \Pi_L^\delta(q^2) = i\int dx e^{iqx} \langle 0 |Tj_\mu^\delta(x)
   j_\nu^\delta(0)|0\rangle \\
   &\mbox{for} \quad \delta=v,a, \nonumber \\
   &q^2 \Pi^\delta(q^2) = i \int dx e^{iqx} \langle 0 |Tj^\delta(x)
   j^\delta(0)|0\rangle \\
   &\mbox{for} \quad \delta=s,p, \nonumber
 \end{align}
with the currents 
\begin{align}
  j_\mu^v = \bar\psi \gamma_\mu \psi, \quad j_\mu^a = \bar\psi
  \gamma_\mu \gamma_5 \psi, \quad j^s = \bar \psi\psi, \quad j^p =
  i \bar\psi\gamma_5 \psi. 
\end{align}
In the low energy limit each of the polarization functions can be written as a
series in $z=\frac{q^2}{4 m^2}$, where $m$ is the mass of the heavy
quark and $Q_q$ the corresponding charge,
\begin{equation}
  \Pi^\delta(q^2) = \frac{3Q_q ^ 2}{16\pi^2} \sum_{n>0} C^\delta_n z^n .
\end{equation}
The coefficients $C^\delta_n$ can be expanded in a power series in
$\frac{\alpha_s}{\pi}$:
\begin{equation}
  C^\delta_n = C^{(0),\delta}_n + \frac{\alpha_s}{\pi} C_F
  C^{(1),\delta}_n + \left (\frac{\alpha_s}{\pi} \right )^2
  C^{(2),\delta}_n + \left (\frac{\alpha_s}{\pi} \right )^3
  C^{(3),\delta}_n + \cdots \,  .
\end{equation}
The decomposition of the four-loop contribution $C^{(3),\delta}_n$  according to the
number of internal quark loops and its colour structure leads to
\begin{align}
\label{eq:9} 
C_n^{(3),\delta} &=\, C_F T_F^2 n_l^2 C_{ll,n}^{(3),\delta} 
+ C_F T_F^2 n_h^2 C_{hh,n}^{(3),\delta}
+ C_F T_F^2 n_l n_h C_{lh,n}^{(3),\delta} +  C_{n_f^0,n}^{(3),\delta} 
\nonumber\\& +  C_F T_F n_l\left ( C_A C_{lNA,n}^{(3),\delta} + C_F
   C_{lA,n}^{(3),\delta} \right )
 +  C_F T_F n_h\left( C_A C_{hNA,n}^{(3),\delta} + C_F
   C_{hA,n}^{(3),\delta} \right )
  \, .
\end{align}
Here $C_F=\frac{N_C^2 - 1}{2N_C}$ and $C_A = N_C$ are the Casimir operators
of the fundamental and adjoint representation of the $SU(N_C)$ group, respectively. 
$T_F=\frac{1}{2}$ is the index of the fundamental representation.
$n_h$ and $n_l$ denote the number of heavy and light quarks,
respectively. $C_{n_f^0,n}^{(3),\delta}$ contains the purely bosonic contributions,
where we set the number of colours $N_C=3$ for simplicity. 

%
As described in \cite{Kuhn:2007vp} the theoretically computed moments $C_n$ can be
combined with measurements of the cross section of $e^+e^- \to \text{hadrons}$ to
determine the masses of charm and bottom quarks. The method is based on 
the dispersion relation
\begin{equation}
  \label{eq:disp}
  \Pi(q^2) = \frac{1}{12\pi^2} \int_0^\infty ds \frac{R(s)}{s(s-q^2)} \, ,
\end{equation}
where
\begin{equation}
  \label{eq:R}
  R(s) = \frac{\sigma(e^+e^- \to \text{hadrons})}{\sigma(e^+e^- \to \mu^+\mu^-)}\,.
\end{equation}
Taylor expanding both sides of Eq. (\ref{eq:disp}) leads to
\begin{equation}
  \label{eq:mdet}
  m=\frac{1}{2}\left(\frac{9 Q_q^2}{4
          }\frac{C_n}{{\cal M}_n^{exp}}\right)^\frac{1}{2n}
\end{equation}
with the experimental moments
\begin{equation}
  \label{eq:Mexp}
  {\cal M}_n^{exp} = \int ds \frac{R(s)}{s^{n+1}}\,.
\end{equation}

%
The calculation of the theoretical moments $C_n$ proceeds as follows:
The diagrams are generated using {\tt qgraf}\cite{Nogueira:1991ex}. In
total there are 701 four-loop diagrams of the propagator type.  Some of
these diagrams are singlet diagrams, i.e. diagrams with massless cuts,
and are not considered in this work. Subsequently the diagrams are
expanded in the external momentum $q^2$ and mapped to six topologies of
vacuum integrals. This is done using {\tt q2e} and {\tt exp} \cite{exp}
in combination with {\sc Matad} \cite{Steinhauser:2000ry} written in
{\tt FORM} \cite{Vermaseren:2000nd}. This procedure leads to integrals
with a maximum of 12 additional powers of propagators and 8 irreducible
scalar products. The large number of integrals obtained in this step can
be reduced to a small set of master integrals solving a large system of
linear equations generated by Integration-by-Parts (IBP) identities
\cite{Chetyrkin:1981qh}. This reduction is achieved using Laporta's
algorithm \cite{Laporta:2001dd} implemented in {\sc Crusher}
\cite{crusher}. This standard method can be assisted by a special
treatment of integrals containing self energies, which is described in
detail in Section \ref{sec:reduct-techn-integr}. It has to be noted that
this aforementioned reduction comprises the most difficult part of the
calculation.  The master integrals have been calculated in
\cite{Chetyrkin:2006dh,Schroder:2005hy,Laporta:2002pg,Chetyrkin:2004fq,Kniehl:2005yc,Schroder:2005db,Bejdakic:2006vg,Kniehl:2006bf,Kniehl:2006bg}. After
performing the renormalization of the quark masses and the strong
coupling constant in the $\overline{\mathrm{MS}}$ scheme the results
given in Section \ref{sec:results} are obtained.


\section{Reduction technique for integrals with internal self energies}
\label{sec:reduct-techn-integr}

\subsection{Algorithm}

The required CPU time for the reduction of Feynman integrals of a given topology to master integrals using Laporta's algorithm strongly depends on the powers of the propagators and scalar products. For a fixed depth of a Taylor expansion in an external momentum of a Feynman amplitude the maximal number of propagator powers will appear on those integrals which contain the maximal number of self energy insertions. In the case of $4$-loop tadpoles we have at most $3$ self energy subgraphs which lead to a sum of propagator powers raised by $2$ compared to the generic case without self energies. Such integrals are thus particularly difficult for Laporta's algorithm. On the other hand the presence of self energy insertions can be exploited to perform the reduction to master integrals in two less expensive steps, first, the reduction of the self energy subgraphs and, second, the reduction of the remaining integral. In the second step Intgration-by-Parts relations are constructed in which the self energy master integrals are treated as objects depending only on their external momentum $q$. In the following the procedure is illustrated for an arbitrary integral which contains a self energy insertion:
\begin{align} \label{initial-se-topo}
  T~=~~\GenericSeTopo~~~~=~~\int d^dq R^{\mu_1\dots \mu_n}(q,m) S_{\mu_1\dots \mu_n}(q,m) P_{m_0}^{a_0}(q^2)\ ,
\end{align}
where $S_{\mu_1\dots \mu_n}(q,m)$ is a rank $n$ tensor self energy integral, $R^{\mu_1\dots \mu_n}(q,m)$ is the rest graph, and $P_{m_0}^{a_0}(q^2) = (q^2-m_0^2)^{-a_0}$ is the connecting propagator with mass $m_0$. All masses are assumed to be either zero or $m$. $S_{\mu_1\dots \mu_n}(q,m)$ is expressed in terms of a complete set of Lorentz structures $\{\Pi^p_{\mu_1\dots \mu_n}\}$ of rank $n$ consisting of the momentum $q$ and the metric $g^{\mu\nu}$, and the (scalar) self energy master integrals $S_z(q^2,m)$:
\begin{align} \label{se-tensor-to-scalar}
  S_{\mu_1\dots \mu_n}(q,m) = \sum_p \Pi^p_{\mu_1\dots \mu_n} \sum_z c_z^p(d,q^2,m) S_z(q^2,m) \ .
\end{align}
The coefficients $c_z^p(d,q^2,m)$ are rational functions of the space-time dimension $d$, the momentum $q^2$, and the mass $m$. Inserting this decomposition into (\ref{initial-se-topo}) gives
\begin{align}
  T &= \int d^dq \sum_p R^{\mu_1\dots \mu_n}(q,m) \Pi^p_{\mu_1\dots \mu_n}(q) P_{m_0}^{a_0}(q^2) \sum_z c_z^p(d,q^2,m) S_z(q^2,m).
\end{align}
The rest graph tensor is contracted with the Lorentz structures to
\begin{align}
  R^{\mu_1\dots \mu_n}(q,m) \Pi^p_{\mu_1\dots \mu_n}(q) = R^p(q^2,m).
\end{align}
From the structure of the IBP identites we know, that the $q^2$ dependence of the denominators of the coefficients $c_z^p(d,q^2,m)$ factorizes into propagator-like objects $P_k(q^2,m)=(q^2-k^2m^2)^{-1}$, where $k$ takes values from a finite set of integers depending on the type of the self energy. Partial fractioning leads to
\begin{align}\label{coeffpartfrac}
  P_{m_0}^{a_0}(q^2) c_z^p(d,q^2,m) = \sum_k \sum_{r_k} \tilde{c}_z^{p,k,r_k}(d) P_k^{r_k}(q^2,m),
\end{align}
where $P_{m_0}(q^2)$ was written as $P_k(q^2,m)$ with $m_0=km$. By
convention we choose $r_k\ge 0$ for $k\ne 0$ so that the decomposition
(\ref{coeffpartfrac}) is unique. After applying these transformations the integral $T$ takes the form
\begin{align}
  T = \int d^dq \sum_p R^p(q^2,m) \sum_z \sum_k \sum_{r_k} \tilde{c}_z^{p,k,r_k}(d) P_k^{r_k}(q^2,m) S_z(q^2,m),
\end{align}
which can be written more conveniently as
\begin{align}\label{T-to-se-topo}
  T = \sum_k \sum_z \sum_p \sum_{r_k} \tilde{c}_z^{p,k,r_k}(d) T^k_z(\rho^p,r_k,m)
\end{align}
with
\begin{align}\label{reducedtopoTkz}
  T^k_z(\rho^p,r_k,m) = \int d^dq\;R^p(q^2,m) P_k^{r_k}(q^2,m) S_z(q^2,m).
\end{align}
$\rho^p$ denotes the set of propagator powers of the rest graph. That means the initial integral $T$ is expressed as a linear combination of integrals of the type $T^k_z(\rho^p,r_k,m)$ in which the self energy insertions appear only as master integrals and all cross talking momenta between the self energy and the rest graph are removed. Figure \ref{t-to-se-topo-fugure} illustrates eq. (\ref{reducedtopoTkz}) for a four-loop vacuum diagram.

\begin{figure}[t]
  \begin{align*}
    \TopoC{SF}~~~&=~~\sum_{b_i,r_0}\tilde{c}_1^{b_i,0,r_0}(d) \TopoD{SF}{0} ~+~ \sum_{b_i,r_2}\tilde{c}_1^{b_i,2,r_2}(d) \TopoD{WF}{2}\\[3mm]
                &+~~\sum_{b_i,r_0}\tilde{c}_2^{b_i,0,r_0}(d) \TopoE{TF}{0} ~+~ \sum_{b_i,r_2}\tilde{c}_2^{b_i,2,r_2}(d) \TopoE{UF}{2}
  \end{align*}
  \caption{The figure above shows an example of eq. (\ref{T-to-se-topo}) for a four-loop tadpole. Note that on the right hand side besides massless (dashed) lines and lines with mass $m$ (solid) also a line with mass $2m$ (double line) appears which is not present in the initial integral. The next step would be to repeat the procedure for the two-loop self energy.}
  \label{t-to-se-topo-fugure}
\end{figure}

The next step is to construct Integration-by-Parts identities for the integrals $T^k_z$ in which the self energy insertions are treated as objects depending only on their external momenta. The identities have the form
\begin{align}\label{ibp-se}
  0 & = \int d^dq\,\frac{\partial}{\partial k_\mu}\ell_\mu R^p(q^2,m) P_k^{r_k}(q^2,m) S_z(q^2,m)\\
    & = \delta_{k\ell}\,d\,T^k_z(\rho^p,r_k,m) + \int d^dq\,\Big[\ell_\mu\frac{\partial}{\partial k_\mu}\Big]_I R^p(q^2,m) P_k^{r_k}(q^2,m) S_z(q^2,m)\nonumber\\
    & = \delta_{k\ell}\,d\,T^k_z(\rho^p,r_k,m)\nonumber\\
    & + \int d^dq \left(\Big[\ell_\mu\frac{\partial}{\partial k_\mu}\Big]_I R^p(q^2,m) \right) P_k^{r_k}(q^2,m) S_z(q^2,m)\nonumber\\
    & + \int d^dq R^p(q^2,m) \left(\ell_\mu\frac{\partial}{\partial k_\mu} P_k^{r_k}(q^2,m) \right) S_z(q^2,m)\nonumber\\
    & + \int d^dq R^p(q^2,m) P_k^{r_k}(q^2,m) \left(\ell_\mu\frac{\partial}{\partial k_\mu} S_z(q^2,m) \right),\nonumber
\end{align}
where $k$ is a loop momentum (i.\,e. either $q$ or a loop momentum of $R^p$) and $\ell$ is a loop momentum or, if present, an external momentum. The notation $\ell_\mu[\partial/\partial k_\mu]_I R^p(q^2,m)$ means that the derivative acts on the integrand of $R^p(q^2,m)$. This part and the contribution $\ell_\mu\partial/\partial k_\mu P_k^{r_k}(q^2,m)$ are treated like in traditional IBP. The difference is the treatment of $\partial/\partial k_\mu S_z(q^2,m)$ in the case $k=q$ where the derivative acts on the self energy. The derivative is explicitly performed and the resulting integrals are reduced to the self energy master integrals $S_z$. The coefficients appearing in this reduction require partial fractioning in $q^2$ again like in (\ref{coeffpartfrac}). Afterwards all terms in (\ref{ibp-se}) are expressed by $T^k_z$ and the system of equations is solved by the Laporta algorithm.

\begin{figure}[t]
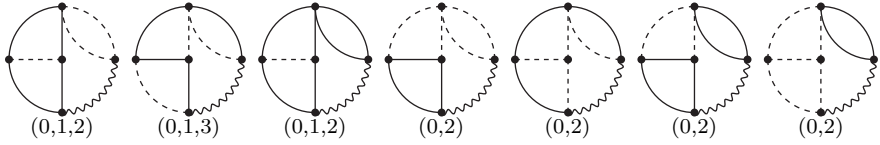

  \begin{center}
    $\TopoB{SA}{(0,1,2)}~~\TopoB{SB}{(0,1,3)}~~\TopoB{SC}{(0,1,2)}~~\TopoB{SD}{(0,2)}~~\TopoB{SE}{(0,2)}~~\TopoB{SF}{(0,2)}~~\TopoB{SG}{(0,2)}$
  \end{center}
  \caption{All four-loop tadpoles with self energy insertions which appear in our calculation can be mapped to these seven topologies. Dotted lines are massless and solid lines carry the mass $m$. The wavy propagator which connects the self energies appears with the masses $km$ where $k$ is an element of the list below the corresponding diagram.}
  \label{se-topos}
\end{figure}

In the case of four-loop tadpoles with a one-loop self energy insertion the restgraph is a two-loop self energy. Therefore the procedure above is applied to the two-loop self energy in a second pass. The topologies which appear in our calculation are depicted in Fig. (\ref{se-topos}).

Let us add a few remarks at this point. The number of integers on which a topology $T^k_z$ depends is given by the number of propagators of the initial topology $T$ minus the number of propagators of the self energy subgraphs and the cross talking scalar products. In the case of four-loop tadpoles only one propagator power of initially ten (eight propagators and two irreducible scalar products) survives this procedure. Therefore the combinatorics of the IBP system for the $T^k_z$ is much better behaved than for the initial topology. There is a price to pay for this simplification, of course. The reduction of subgraphs and the partial fractioning of propagators which carry the same momentum but different masses leads to a system of IBP relations which couples different topologies. Reducing a single propagator power from a self energy can produce two powers on an external leg, therefore in the worst case the total number of propagator powers might be doubled. If the rest graph is sufficiently simple these issues are easy to deal with. Futhermore, the tensor reduction of large powers of cross-talking scalar products will produce huge intermediate expressions and many terms in eq. (\ref{T-to-se-topo}). These expressions simplify to the full reduction to master integrals when the solution of the IBP system (\ref{ibp-se}) is inserted. The procedure was implemented in a \texttt{Mathematica} program.

\subsection{Implementation in {\tt FIRE}}

Additionally to the stand-alone implementation mentioned above, for
one-loop self energy insertions the procedure was also realized with the
help of the {\tt FIRE}  algorithm \cite{Smirnov:2008iw}.

Instead of the tensor reduction (\ref{se-tensor-to-scalar}) the
so-called \textit{region-bases} feature of \texttt{FIRE} is used. The
tensor structure arises from scalar products of loop-momenta between the
self energy and the rest graph. The \texttt{SBases} program
\cite{Smirnov:2006tz} can construct a basis of recursion relations to
reduce the power of these cross-talking scalar products to zero without
caring about the rest of the intgeral. Such a basis is called a
\textit{region-basis}. In the second step the scalar self energy
integrals are reduced to master integrals. This can either be done by a
region-basis, or with the help of precalculated reduction tables in
combination with the \textit{Rules} feature of \texttt{FIRE}.

In the case of one-loop self energies with one mass (either massive-massive or massive-massless) two master integrals appear, the massive tadpole and the self energy with both propagator powers equal to one. Integrals which contain the tadpole are treated like usual integrals with one loop less than the initial integral, because the tadpole factorizes from the rest of the intgeral. The treatment of integrals with the self energy master insertion is a bit more complicated. One has to introduce a new object in Feynman diagrams. We call this a \textit{heavy dot} which can be either present (integral has a self energy master insertion), or not (integral has a tadpole insertion) and is therefore represented by an index which is either zero or one.

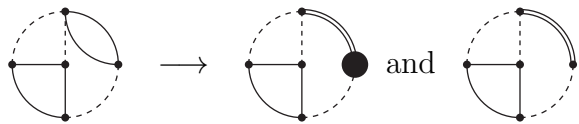
\begin{center}
  $\TopoA{SFo}\quad\longrightarrow\quad\FireSeTopo{1}\quad$and$\quad\FireSeTopo{0}$
\end{center}

Note that we also have to introduce additional lines which carry masses depending on the self energy type. The massive-massive self energy needs a massless and a double-massive line, the massive-massless self energy needs a massless and a usual massive line that is external with respect to the self energy (one of these is naturally present in the initial diagram). These lines are needed to absorb the momentum dependence of the coefficients from the reduction of the self energies to master integrals. The partial fractioning of these two lines is done by a recursion relation which is added to the region basis. The IPBs for the remaining integrals are constructed as in eq. (\ref{ibp-se}) where the creation or destruction of a heavy dot is done by a shift operator for the corresponding index, just like for usual propagators. The reduction procedure aims on reducing the heavy dot index to zero if possible. The {\tt FIRE} program has a setting allowing to use the heavy dots and can create SBases for integrals with heavy dots present. However, the IBP generator coming with \texttt{FIRE} at the moment is not able to construct such IBP relation. As it has been explained above, the procedure can be repeated if a second self energy is present and one can again consider a region-basis corresponding to the second self energy subdiagram, insert rules reducing it to masters and finally result in a problem with two heavy dots.

\subsection{Top level reduction}

Although the approach from above is not applicable to general Feynman
integrals, in combination with \texttt{FIRE} for the top level reduction
it can be used for the calculation of arbitrary integrals. In the
highest sectors the SBases are used to reduce the integrals until enough
lines are contracted so that only integrals with internal self energies
are left. From experience we know that the SBases algorithm often fails
in sectors where self energy insertions are present. At this point a
program that implements the above procedure is plugged in with the rules
framework of \texttt{FIRE} to automatically finish the reduction. This
way we do not need SBases for the lower sectors for which in some cases
we could not find a basis. In combination with the top level reduction
single integrals can be calculated if needed or for checks. However, at
present this approach is not suited for large scale calculations. The
self energy formalism itself is very efficient and can be used to
significantly reduce the effort to invest in Laporta algorithm.



\section{Results}
\label{sec:results}
In this Section we present the numerical values of all moments known at
${\cal O}(\alpha_s^3)$ for the different currents. The full analytical
expressions of the new four-loop contributions can be found in Appendix
\ref{sec:analytical-results}.  We present  results up to the third
moment for the scalar, vector and axial-vector correlators. In the case of the
pseudo-scalar current the $n$-th moment is related to the $(n-1)$-th moment
of the longitudinal part of the axial-vector current through the Ward
identity
\begin{equation}
  q^2 \Pi_L^a(q^2) = 4m^2 (\Pi^p(q^2) - q^2 (\partial \Pi^p (q^2) /
  \partial q^2 ) |_{q^2 = 0 }) \, .
\end{equation}
This allows to obtain the fourth moment of the pseudo-scalar current. 
The numerical values are
obtained setting the renormalization scale $\mu = m$ and using the
$\overline{\mathrm {MS}}$ scheme for the renormalization of the quark
masses. We explicitely keep the dependence on the number of light quarks
$n_l$ and set the number of heavy quarks $n_h=1$. For completeness we
give the numerical values for all terms in the perturbative series in
$\alpha_s$ up to ${\cal O}(\alpha_s^3)$.
\begin{align*}
C_1^v=&1.06666 + 2.55473\frac{\alpha_s}{\pi} + 
   (0.50988 + 0.66227n_l) \left(\frac{\alpha_s}{\pi}\right)^2\\
   &+ 
   (1.87882 - 2.79472n_l +      0.09610n_l^2) \left(\frac{\alpha_s}{\pi}\right)^3,\\
C_2^v=&0.45714 + 1.10955\frac{\alpha_s}{\pi} + 
    (1.41227 + 0.45491n_l)   \left(\frac{\alpha_s}{\pi}\right)^2\\
   &+ 
    (-6.23488 + 0.96156n_l - 0.01594n_l^2)   \left(\frac{\alpha_s}{\pi}\right)^3,\\ 
C_3^v=&0.27089 + 0.51939\frac{\alpha_s}{\pi} + 
    (0.35222 + 0.42886n_l)   \left(\frac{\alpha_s}{\pi}\right)^2\\
   &+ 
    (-8.30971 + 1.94219n_l - 0.03959n_l^2)   \left(\frac{\alpha_s}{\pi}\right)^3\\
  \end{align*}  
\begin{align*}
C_1^a=&0.53333 + 0.84609\frac{\alpha_s}{\pi} + 
   (-2.34665 + 0.41316n_l) \left(\frac{\alpha_s}{\pi}\right)^2\\
   &+ 
   (-1.16280 - 0.56583n_l + 
     0.04784n_l^2) \left ( \frac{\alpha_s}{\pi}\right ) ^3,\\ 
C_2^a=&0.15238 + 0.14165\frac{\alpha_s}{\pi} + 
    (-0.83002 + 0.19218n_l)   \left(\frac{\alpha_s}{\pi}\right)^2\\
   &+ 
    (-6.95414 + 1.11092n_l - 0.02049n_l^2)   \left(\frac{\alpha_s}{\pi}\right)^3,\\ 
C_3^a=&0.06772 - 0.01276\frac{\alpha_s}{\pi} + 
    (-0.67592 + 0.13562n_l)   \left(\frac{\alpha_s}{\pi}\right)^2\\
   &+ 
    (-5.36382 + 0.98605n_l - 0.02233n_l^2)   \left(\frac{\alpha_s}{\pi}\right)^3,\\
  \end{align*}
\begin{align*}
C_1^s=&0.8 + 0.60246\frac{\alpha_s}{\pi} + 
 (-9.50321 + 0.58765n_l) \left(\frac{\alpha_s}{\pi}\right)^2\\
   &+ 
 (2.36044 - 3.31076n_l +    0.23981n_l^2) \left(\frac{\alpha_s}{\pi}\right)^3,\\
C_2^s=&0.22857 + 0.42582\frac{\alpha_s}{\pi} + 
    (-1.44346 + 0.23664n_l)   \left(\frac{\alpha_s}{\pi}\right)^2\\
   &+ 
    (-16.84601 + 1.28345n_l + 0.00398n_l^2)   \left(\frac{\alpha_s}{\pi}\right)^3,\\ 
C_3^s=&0.10158 + 0.15355\frac{\alpha_s}{\pi} + 
    (-0.60909 + 0.15633n_l)   \left(\frac{\alpha_s}{\pi}\right)^2\\
   &+ 
    (-10.40154 + 1.32090n_l - 0.01802n_l^2)
    \left(\frac{\alpha_s}{\pi}\right)^3,\\
  \end{align*}
\begin{align*}
  C_1^p=&1.33333 + 3.11111\frac{\alpha_s}{\pi} + (-1.73650 + 0.61728 n_l)\left(\frac{\alpha_s}{\pi}\right)^2\\
& +    (21.34792 - 8.66336n_l + 0.37997n_l^2)\left(\frac{\alpha_s}{\pi}\right)^3\\
C_2^p=&0.53333 + 2.06419\frac{\alpha_s}{\pi} + 
  (6.36704 + 0.28971n_l) \left(\frac{\alpha_s}{\pi}\right)^2\\
   &+ 
  (10.92473 - 1.49687n_l + 
    0.07020n_l^2)\left ( \frac{\alpha_s}{\pi}\right ) ^3,\\
C_3^p=&0.30476 + 1.21171\frac{\alpha_s}{\pi} + 
    (5.19573 + 0.26782n_l)   \left(\frac{\alpha_s}{\pi}\right)^2\\
   &+ 
    (13.96839 + 0.15741n_l + 0.01535n_l^2)   \left(\frac{\alpha_s}{\pi}\right)^3,\\ 
C_4^p=&0.20317 + 0.71275\frac{\alpha_s}{\pi} + 
    (3.40816 + 0.28627n_l)   \left(\frac{\alpha_s}{\pi}\right)^2\\
   &+ 
    (10.20740 + 1.06763n_l - 0.00916n_l^2)
    \left(\frac{\alpha_s}{\pi}\right)^3,\\
  \end{align*}
  We find full analytical and numerical agreement with all previously
  known results, which means  $C_1^v$ from Refs
  \cite{Chetyrkin:2006xg,Boughezal:2006px} and $C_1^\delta$ and $C_2^p$ from
  Ref. \cite{Sturm:2008eb}.

  In \cite{Hoang:2008qy} the value of the third moment of the vector
  current has been predicted $C_3^v = -3.279 \pm 0.573 , -1.457 \pm
  0.579$ for $n_l=3,4$, respectively. The calculated values $-2.839$ and
  $-1.174$ are within these error bound. They also lie well within the
  estimates of  \cite{Kuhn:2007vp}.

\section{Applications}
\label{sec:applications}

One of the main applications of the low-energy moments of the vacuum
polarization function is the determination of the charm- and
bottom-quark masses using sum rules. As can be seen from
Eq. (\ref{eq:mdet}) the masses can be extracted from the experimental
measurement of the $R$-ratio in the threshold region of open charm or
bottom production. This relation has been exploited in
\cite{Kuhn:2007vp} where only the first moment of the vector correlator
was used. Recently, new data from the Babar collaboration was published
for the region around the $b\bar b$ threshold \cite{:2008hx}. Since a
full analysis of this new data is beyond the scope of this paper, we
only quote the results from \cite{Chet:2009}, where besides the new
Babar data also all moments presented in this work were included. 
It turns out that our new results lead to a reduction
of the theoretical uncertainty by about 6 -- 10 MeV while the mean
values extracted using different moments get
only shifted by about 2 -- 4 MeV. The values of the quark masses obtained
from the different moments are in good agreement which demonstrates the
consistency of the method. 
 The final values for the charm and bottom quark read
\begin{align}
  m_c(3\mbox{GeV}) &= 0.986(13)\mbox{GeV} \, ,\\ 
  m_b(m_b) &= 4.163(16) \mbox{GeV} \, .
\end{align}

In the case of the charm quark the experimental data can be replaced by
lattice ``data'', i.e. data obtained from the calculation of appropriate
correlators on the lattice. This has recently been done by the HPQCD
collaboration in Ref.  \cite{Allison:2008xk}. They found that in
practice the pseudo-scalar correlator is best suited for these
calculations, but also the other currents can be used with less
accuracy. In Table \ref{tab:charm} we give an update of the results in
\cite{Allison:2008xk} where our result for the third moment of the
pseudo-scalar correlator was already used prior to this
publication.  The
values of the charm-quark mass extracted using different moments are in
very good agreement with each other and the results from analyses using
experimental data as input.
 \begin{table}
 \centering
 \begin{tabular}{c|c|c}
 moment& $m_c$(3 GeV)[GeV] Ref. \cite{Allison:2008xk}&$m_c$(3 GeV)[GeV] \\\hline\hline
 2     &0.986(11)                                    &  0.986(11)       \\\hline
 3     &0.986(10)                                    &0.986(10)         \\\hline
 4     &0.973(19)                                    &0.981(13)         \\\hline
 5     &0.969(23)                                    & 0.975(17)        
 \end{tabular}
 \caption{Update of the results from Table II given in \cite{Allison:2008xk} for $m_c$
   extracted using the pseudo-scalar correlator. The value of the fifth
   moment was taken from Ref. \cite{Kiyo:2009}.}
\label{tab:charm}
 \end{table}

Lattice calculations can also be used to determine the value of the
strong coupling constant as explained in detail in
\cite{Allison:2008xk}. Again the pseudo-scalar correlator is best suited
and in Table \ref{tab:alphas} we give an update of
the numbers presented in \cite{Allison:2008xk}.  The values from
different moments are again in very good agreement with each other and
competitive with the world average for $\alpha_s(M_Z) = 0.1176(20)$
\begin{table}
  \centering
  \begin{tabular}{c|c|c|c}
moment&$\alpha_s(3 \,\mbox{GeV})$  Ref. \cite{Allison:2008xk}&$\alpha_s(3 \, \mbox{GeV})$ & $\alpha_s(M_Z)$ \\\hline\hline
1     & 0.252(6) &0.252(6)                    & 0.1177(12)   \\\hline
2/3   & 0.249(6) &0.249(6)                    & 0.1170(13)   \\\hline
3/4   & 0.224(31)&0.237(11)                   & 0.1145(25)   \\\hline
4/5   & 0.241(30)&0.236(19)                   & 0.1143(44)
  \end{tabular}
  \caption{Update of the results for $\alpha_s$ from Table II given in
    \cite{Allison:2008xk}. The values of $\alpha_s$ are extracted using
    the indicated ratios of moments. The value of the fifth
   moment was taken from Ref. \cite{Kiyo:2009}.}
\label{tab:alphas}
\end{table}
\section{Conclusion}
\label{sec:conclusion}
We calculated the second and third low-energy moment of the heavy quark
correlator for vector, scalar and axial-vector currents and derived the
fourth moment of the pseudo-scalar correlator.  To this end we discussed
new methods based on Sbases and reduction of internal self-energies
for the reduction of scalar integrals to masters.  Furthermore we give
an update of the results for the quark masses and $\alpha_s$ obtained in
previous publications. The new results presented reduce the uncertainty on the
quark masses by about 6 -- 10 MeV while the central  values are only
shifted by about 2 -- 4 MeV. This demonstrates the validity of the method
used for the mass determination. Furthermore the new results can be used
to improve the reconstruction of $\Pi(q^2)$ over the whole energy range
using Pad\'e approximation.

\section*{Acknowledgements}\label{sec:acknowledgements}

We thank K.\,G.~Chetyrkin and J.\,H.~K\"uhn for helpful discussions and
M.~Steinhauser and C.~Sturm for providing the update of the  lattice results. 
This work was supported by the Deutsche Forschungsgemeinschaft through
the SFB/TR-9 ``Computational Particle Physics''. Ph.\,M. was supported
by the Gra\-du\-ier\-ten\-kolleg ``Hochenergiephysik und Teilchenastrophysik''.
A.\,M. thanks the Landesgraduiertenf\"orderung for support. The work of
A.\,S. was supported in part by the Russian Foundation for
Basic Research through grant 08-02-01451.

\nocite{*}

\begin{appendix}
\section{Analytical Results}
\label{sec:analytical-results}

In this Appendix we list the analytical results for the different
currents where we used the abbreviations
\begin{align}
c_4=&24a_4+\log^4(2)-6\zeta_2\log^2(2)\,,\notag\\
a_n=&\mathrm{Li}_n(\tfrac{1}{2}) = \sum_{k=1} ^ \infty \frac{1}{2^k k^n} \,,\notag\\
\zeta_n =& \sum_{k=1}^\infty \frac{1}{k^n}\, . 
\end{align}
The results are split according to the colour structures defined in
Eq. (\ref{eq:9}) and use the $\overline { \mathrm{MS} }$-scheme for the
renormalization of the quark masses. 

The results of this calculation can be downloaded  in computer readable form
from \\\texttt{ http://www-ttp.particle.uni-karlsruhe.de/Progdata/ttp09/ttp09-18/}.

\subsection{Second moments}
\label{sec:second-moments}

\subsubsection{Vector current}
\label{sec:vector-current}

\begin{align*}
  C^{(3),v}_{ll,2}=&\frac{15441973}{19136250}-\frac{32}{45}\zeta_3,\\
 C^{(3),v}_{hh,2}=&\frac{1842464707}{646652160}-\frac{2744471}{1064448}\zeta_3,\\
 C^{(3),v}_{lh,2}=&\frac{95040709}{62705664}-\frac{2029}{41472}c_4+\frac{99421}{55296}\zeta_4-\frac{12159109}{4644864}\zeta_3,\\
 C^{(3),v}_{lNA,2}=&-\frac{22559166733}{16796160000}-\frac{520999}{4354560}c_4+\frac{167529079}{5806080}\zeta_4-\frac{309132631}{12902400}\zeta_3,\\
 C^{(3),v}_{lA,2}=&\frac{357543003871}{11757312000}+\frac{520999}{2177280}c_4+\frac{598455689}{2903040}\zeta_4-\frac{36896356307}{174182400}\zeta_3,\\
 C^{(3),v}_{hNA,2}=&-\frac{20427854209619}{5649153269760}-\frac{31595849}{11612160}c_4+\frac{968787977}{15482880}\zeta_4+\frac{362}{63}\zeta_5\\
 & -\frac{29638030087837}{697426329600}\zeta_3,\\
 C^{(3),v}_{hA,2}=&-\frac{37320009196157}{271593907200}-\frac{130387543}{2177280}c_4+\frac{2218910663}{1451520}\zeta_4\\
& -\frac{5811074101069}{6706022400}\zeta_3,\\
 C^{(3),v}_{n_f^0,2}=&\frac{64985074258811347}{353072079360000}-\frac{1662518706713}{21016195200}c_4-\frac{26401638588211}{28021593600}\zeta_4\\
 &
 -\frac{164928917}{270270}\zeta_5-\frac{2900811008}{3648645}a_5-\frac{1684950406}{3648645}\log(2)\zeta_4\\
&-\frac{725202752}{10945935}\log^3(2)\zeta_2
  +\frac{362601376}{54729675}\log^5(2)\\
& +\frac{112680551036302633}{47076277248000}\zeta_3 \, .
\end{align*}

\subsubsection{Scalar current}
\begin{align*}
  C^{(3),s}_{n_f,2}=&\,\frac{381690470169079}{467026560000}+\frac{66906848}{1216215}a_5+\frac{402167880157}{28021593600}c_4\\
  & -\frac{8363356}{18243225}\log^5(2)+\frac{16726712}{3648645}\log^3(2)\zeta_2-\frac{87227759}{1216215}\log(2)\zeta_4\\
  & -\frac{1075951565336201}{560431872000}\zeta_3+\frac{3294508921817}{5337446400}\zeta_4+\frac{480273679}{694980}\zeta_5,\displaybreak[0]\\
  C^{(3),s}_{hN,2}=&\,-\frac{355706449933319}{37661021798400}-\frac{4135403}{2150400}c_4-\frac{857749280491}{37196070912}\zeta_3\\
  & +\frac{83367701}{1720320}\zeta_4-\frac{25}{21}\zeta_5,\displaybreak[0]\\
  C^{(3),s}_{lN,2}=&\,-\frac{102230727187}{19595520000}-\frac{1121}{48384}c_4+\frac{986734811}{58060800}\zeta_3-\frac{4419371}{322560}\zeta_4,\displaybreak[0]\\
  C^{(3),s}_{hA,2}=&\,\frac{589278012260869}{2353813862400}+\frac{21022661}{181440}c_4+\frac{98124313871713}{58118860800}\zeta_3,\displaybreak[0]\\
  & -\frac{357046549}{120960}\zeta_4\\
  C^{(3),s}_{lA,2}=&\,-\frac{12729644591}{279936000}+\frac{1121}{24192}c_4+\frac{1192980407}{9676800}\zeta_3-\frac{5073991}{53760}\zeta_4,\displaybreak[0]\\
  C^{(3),s}_{lh,2}=&\,\frac{64429199}{41803776}-\frac{1301}{27648}c_4-\frac{37067899}{15482880}\zeta_3+\frac{63749}{36864}\zeta_4,\displaybreak[0]\\
  C^{(3),s}_{hh,2}=&\,\frac{34805101}{6735960}-\frac{101467}{23760}\zeta_3,\displaybreak[0]\\
  C^{(3),s}_{ll,2}=&\,\frac{1439317}{6378750}-\frac{8}{45}\zeta_3  \, .
\end{align*}

\subsubsection{Axial-vector current}

\begin{align*}
  C^{(3),a}_{n_f,2}=&\,\frac{217627345572539363}{29957630976000}-\frac{22929814528}{18243225}a_5+\frac{48366697030607}{367783416000}c_4\\
  &
  +\frac{2866226816}{273648375}\log^5(2)-\frac{5732453632}{54729675}\log^3(2)\zeta_2
  \\ 
& -\frac{5480723618}{2606175}\log(2)\zeta_4
   -\frac{19481983397047414063}{3295339407360000}\zeta_3\\
& -\frac{3999530651550337}{980755776000}\zeta_4+\frac{64302085207}{12162150}\zeta_5,\\
  C^{(3),a}_{hN,2}=&\,-\frac{58553635489580981}{3954407288832000}-\frac{5032917199}{3048192000}c_4-\frac{289668262443149}{19527937228800}\zeta_3\\
  & +\frac{40203350213}{812851200}\zeta_4-\frac{382}{45}\zeta_5,\displaybreak[0]\\
  C^{(3),a}_{lN,2}=&\,-\frac{414223585133}{33592320000}-\frac{2041307}{43545600}c_4+\frac{43700406067}{696729600}\zeta_3-\frac{95269223}{1658880}\zeta_4,\displaybreak[0]\\
  C^{(3),a}_{hA,2}=&\,\frac{30093199294856077}{35307207936000}+\frac{707672969}{1814400}c_4+\frac{1650262562573011}{290594304000}\zeta_3\\
  & -\frac{1718278879}{172800}\zeta_4,\displaybreak[0]\\
  C^{(3),a}_{lA,2}=&\,-\frac{4855444846081}{23514624000}+\frac{2041307}{21772800}c_4+\frac{30349640587}{49766400}\zeta_3-\frac{2828510767}{5806080}\zeta_4,\displaybreak[0]\\
  C^{(3),a}_{lh,2}=&\,\frac{809030351}{627056640}-\frac{2797}{82944}c_4-\frac{92264351}{46448640}\zeta_3+\frac{137053}{110592}\zeta_4,\displaybreak[0]\\
  C^{(3),a}_{hh,2}=&\,\frac{1173822473}{323326080}-\frac{24710027}{7983360}\zeta_3,\displaybreak[0]\\
  C^{(3),a}_{ll,2}=&\,\frac{4275743}{19136250}-\frac{32}{135}\zeta_3  \, .
\end{align*}

\subsection{Third moments}
\label{sec:third-moments}

\subsubsection{Vector current}
\label{sec:vector-current-1}

\begin{align*}
  C^{(3),v}_{ll,3}=&\frac{31556642272}{49228003125}-\frac{256}{405}\zeta_3,\\
 C^{(3),v}_{hh,3}=&\frac{56877138427}{12609717120}-\frac{6184964549}{1556755200}\zeta_3,\\
 C^{(3),v}_{lh,3}=&\frac{60361465477}{29393280000}-\frac{1765}{31104}c_4+\frac{86485}{41472}\zeta_4-\frac{57669161}{17418240}\zeta_3,\\
 C^{(3),v}_{lNA,3}=&-\frac{1475149211788337}{6452412825600000}-\frac{8529817}{77414400}c_4+\frac{1510937903}{14745600}\zeta_4\\
 & -\frac{561258009401}{6193152000}\zeta_3,\\
 C^{(3),v}_{lA,3}=&\frac{983812946922223}{4389396480000}+\frac{8529817}{38707200}c_4+\frac{21972351293}{17203200}\zeta_4\\
 & -\frac{28995540810097}{21676032000}\zeta_3,\\
 C^{(3),v}_{hNA,3}=&-\frac{454880458419083629}{5854170457175040000}-\frac{7110196837}{1117670400}c_4+\frac{1068488091383}{7451136000}\zeta_4\\
 & +\frac{4448}{315}\zeta_5-\frac{43875740175477222611}{433642256087040000}\zeta_3,\\
 C^{(3),v}_{hA,3}=&-\frac{2327115263308753}{2489610816000}-\frac{16870125343}{39916800}c_4+\frac{286864384271}{26611200}\zeta_4\\
 & -\frac{377837317054807}{61471872000}\zeta_3,\\
 C^{(3),v}_{n_f^0,3}=&\frac{8011001677156303009183663}{1270551061901721600000}-\frac{16091704629458603}{45731240755200}c_4\\
 & -\frac{1505000915688143609}{304874938368000}\zeta_4-\frac{1781851011826}{310134825}\zeta_5-\frac{859399602944}{310134825}a_5\\
 &
 -\frac{38830116184}{44304975}\log(2)\zeta_4-\frac{214849900736}{930404475}\log^3(2)\zeta_2
 \\ & +\frac{107424950368}{4652022375}\log(2)^5
  +\frac{1061162538194750079871}{128047474114560000}\zeta_3  \, .
\end{align*}

\subsubsection{Scalar current}
\label{sec:scalar-current}

\begin{align*}
  C^{(3),s}_{ll,3}=&\frac{2229649183}{16409334375}-\frac{64}{405}\zeta_3,\\
 C^{(3),s}_{hh,3}=&\frac{82738947097}{12009254400}-\frac{11973270941}{2075673600}\zeta_3,\\
 C^{(3),s}_{lh,3}=&\frac{1047601560409}{627056640000}-\frac{139777}{3317760}c_4+\frac{6849073}{4423680}\zeta_4-\frac{4641507553}{1857945600}\zeta_3,\\
 C^{(3),s}_{lNA,3}=&-\frac{22504447164310009}{1075402137600000}-\frac{2832217}{116121600}c_4-\frac{288902173}{4423680}\zeta_4\\
 & +\frac{142693634059}{1857945600}\zeta_3,\\
 C^{(3),s}_{lA,3}=&-\frac{296129104756579}{1024192512000}+\frac{2832217}{58060800}c_4-\frac{51185293321}{77414400}\zeta_4
 \displaybreak[0]\\ \displaybreak[0] &  +\frac{1811694475921}{2167603200}\zeta_3,\\
 C^{(3),s}_{hNA,3}=&-\frac{549859189388336126083}{31222242438266880000}-\frac{270366644899}{89413632000}c_4+\frac{1953161743091}{23843635200}\zeta_4\\
 & -\frac{1888}{315}\zeta_5-\frac{16019821402882223177}{462551739826176000}\zeta_3,\\
 C^{(3),s}_{hA,3}=&\frac{1248837940639034917}{838406080512000}+\frac{108697503829}{159667200}c_4-\frac{263931412979}{15206400}\zeta_4\\
 & +\frac{4308361964639349413}{434729078784000}\zeta_3,\\
 C^{(3),s}_{n_f^0,3}=&\frac{1831719443479347906961831}{211758510316953600000}+\frac{23476362679129111}{266765571072000}c_4\\
 & -\frac{227245780513489691}{71137485619200}\zeta_4+\frac{550981545544}{103378275}\zeta_5-\frac{31233089024}{103378275}a_5\\
 &
 -\frac{132939852688}{103378275}\log(2)\zeta_4-\frac{7808272256}{310134825}\log^3(2)\zeta_2
 \\ &+\frac{3904136128}{1550674125}\log(2)^5
  -\frac{6840706109244803149}{798870159360000}\zeta_3  \, .
\end{align*}

\subsubsection{Axial-vector current}
\label{sec:axial-cevtor-current}

\begin{align*}
  C^{(3),a}_{ll,3}=&\frac{6052378456}{49228003125}-\frac{64}{405}\zeta_3,\\
 C^{(3),a}_{hh,3}=&\frac{599209514131}{126097171200}-\frac{12494549257}{3113510400}\zeta_3,\\
 C^{(3),a}_{lh,3}=&\frac{4962625369889}{3762339840000}-\frac{630217}{19906560}c_4+\frac{30880633}{26542080}\zeta_4-\frac{21887209193}{11147673600}\zeta_3,\\
 C^{(3),a}_{lNA,3}=&-\frac{41502247972857197}{955913011200000}-\frac{254785409}{8360755200}c_4-\frac{2632884019279}{11147673600}\zeta_4\\
 & +\frac{55622383674929}{222953472000}\zeta_3,\\
 C^{(3),a}_{lA,3}=&-\frac{407919130830390283}{368709304320000}+\frac{254785409}{4180377600}c_4-\frac{17645944577969}{5573836800}\zeta_4\\
 & +\frac{2942092239074473}{780337152000}\zeta_3,\\
 C^{(3),a}_{hNA,3}=&-\frac{9291278450921361191597}{266210698684170240000}-\frac{131718995983}{1609445376000}c_4
 \\ & +\frac{12854292041443}{429185433600}\zeta_4
  -\frac{9376}{405}\zeta_5+\frac{29030831574992389907}{1314620734242816000}\zeta_3,\\
 C^{(3),a}_{hA,3}=&\frac{167314115446749909343}{35213055381504000}+\frac{6229030586993}{2874009600}c_4-\frac{15127004909623}{273715200}\zeta_4\\
 & +\frac{49375038643781532767}{1565024683622400}\zeta_3,\\
 C^{(3),a}_{n_f^0,3}=&\frac{1229252060576583400157229437}{22869919114230988800000}+\frac{26069736016065697819}{28810681675776000}c_4\\
 &
 -\frac{84934038473664674083}{1536569689374720}\zeta_4+\frac{29147971741346}{930404475}\zeta_5
 \\ & -\frac{6290348482816}{930404475}a_5
 -\frac{11153783461592}{930404475}\log(2)\zeta_4\\ &-\frac{1572587120704}{2791213425}\log^3(2)\zeta_2
 +\frac{786293560352}{13956067125}\log(2)^5\\ &
  -\frac{18647336379797344185371}{1075598782562304000}\zeta_3  \, . 
\end{align*}

\subsubsection{Pseudo-scalar current}
\label{sec:pseudo-scal-curr-1}

\begin{align*}
  C^{(3),p}_{ll,3}=&\frac{5893303}{9568125}-\frac{64}{135}\zeta_3,\\
 C^{(3),p}_{hh,3}=&-\frac{26179537}{58786560}+\frac{76295}{290304}\zeta_3,\\
 C^{(3),p}_{lh,3}=&\frac{103200341}{627056640}-\frac{1435}{82944}c_4+\frac{70315}{110592}\zeta_4-\frac{5574817}{9289728}\zeta_3,\\
 C^{(3),p}_{lNA,3}=&\frac{102523166819}{11197440000}-\frac{259289}{4838400}c_4+\frac{3760967}{36864}\zeta_4-\frac{927288197}{9289728}\zeta_3,\\
 C^{(3),p}_{lA,3}=&\frac{1867646866703}{7838208000}+\frac{259289}{2419200}c_4+\frac{184831769}{215040}\zeta_4-\frac{16109807909}{16588800}\zeta_3,\\
 C^{(3),p}_{hNA,3}=&\frac{10104443767785889}{1318135762944000}-\frac{4409031071}{677376000}c_4+\frac{25692311767}{180633600}\zeta_4\\
 & +\frac{368}{15}\zeta_5-\frac{248471102645509}{2169770803200}\zeta_3,\\
 C^{(3),p}_{hA,3}=&-\frac{11809333374019073}{11769069312000}-\frac{1657837871}{3628800}c_4+\frac{28182137887}{2419200}\zeta_4\\
 & -\frac{214584455705773}{32288256000}\zeta_3,\\
 C^{(3),p}_{n_f^0,3}=&-\frac{841507212739667387}{549223234560000}-\frac{62193803213161}{245188944000}c_4-\frac{9141288770183}{5108103000}\zeta_4\\
 &
 -\frac{58401651592}{6081075}\zeta_5+\frac{15214899712}{6081075}a_5+\frac{3649219472}{868725}\log(2)\zeta_4\\
 & +\frac{3803724928}{18243225}\log^3(2)\zeta_2
 -\frac{1901862464}{91216125}\log^5(2)\\
 &+\frac{767220341123064149}{84495882240000}\zeta_3  \, .
\end{align*}

\subsection{Fourth Moments}
\label{sec:fourth-moments}

\subsubsection{Pseudo-scalar current}
\label{sec:pseudo-scal-curr}

\begin{align*}
  C^{(3),p}_{ll,4}=&\frac{329624056}{607753125}-\frac{64}{135}\zeta_3,\\
 C^{(3),p}_{hh,4}=&\frac{1739021393}{4670265600}-\frac{52674403}{115315200}\zeta_3,\\
 C^{(3),p}_{lh,4}=&\frac{6245190619}{9289728000}-\frac{18589}{737280}c_4+\frac{910861}{983040}\zeta_4-\frac{1587980983}{1238630400}\zeta_3,\\
 C^{(3),p}_{lNA,4}=&\frac{331245076513169857}{8603217100800000}-\frac{964223}{15925248}c_4+\frac{63056544809}{148635648}\zeta_4\\
 & -\frac{1229566890599}{2972712960}\zeta_3,\\
 C^{(3),p}_{lA,4}=&\frac{1309551728160611}{910393344000}+\frac{964223}{7962624}c_4+\frac{10470510368767}{1857945600}\zeta_4\\
 & -\frac{36245830807891}{5780275200}\zeta_3,\\
 C^{(3),p}_{hNA,4}=&\frac{33124023975801583681957}{1686001091666411520000}-\frac{2883014289719}{107296358400}c_4\\
 & +\frac{87140698714231}{143061811200}\zeta_4 +\frac{11776}{189}\zeta_5-\frac{3691162910334747972379}{8325931316871168000}\zeta_3,\\
 C^{(3),p}_{hA,4}=&-\frac{217480554135087391943}{35213055381504000}-\frac{2695980497779}{958003200}c_4\\
 &+\frac{45831907340963}{638668800}\zeta_4 -\frac{106836033862656915617}{2608374472704000}\zeta_3,\\
 C^{(3),p}_{n_f^0,4}=&\frac{160999779080808137440043699}{7623306371410329600000}-\frac{15601462649447899319}{9603560558592000}c_4\\
 & -\frac{41651006271487731961}{12804747411456000}\zeta_4-\frac{3679557604076}{62026965}\zeta_5+\frac{133358649856}{8860995}a_5\\
 &
 +\frac{1559619599504}{62026965}\log(2)\zeta_4+\frac{33339662464}{26582985}\log^3(2)\zeta_2\\
 & -\frac{16669831232}{132914925}\log(2)^5
 +\frac{519180776457371224187}{21090172207104000}\zeta_3  \, .
\end{align*}

\end{appendix}


\end{document}